# A Novel Compressed Sensing Technique for Traffic Matrix Estimation of Software Defined Cloud Networks

**Sameer Qazi, Syed Muhammad Atif and Muhammad Bilal Kadri**
**College of Engineering,**
PAF Karachi Institute of Economics and Technology (PAF-KIET)
Pakistan
[email: sameer.qazi@pafkiet.edu.pk]
*Corresponding author: Sameer Qazi

## *Abstract*

Traffic Matrix estimation has always caught attention from researchers for better network management and future planning. With the advent of high traffic loads due to Cloud Computing platforms and Software Defined Networking based tunable routing and traffic management algorithms on the Internet, it is more necessary as ever to be able to predict current and future predicted traffic volumes on the network. For large networks such origin-destination traffic prediction problem takes the form of a large under- constrained and under-determined system of equations with a dynamic measurement matrix. Previously, the researchers had relied on the assumption that the measurement (routing) matrix is stationary due to which the schemes are not suitable for modern software defined networks. In this work, we present our Compressed Sensing with Dynamic Model Estimation (CS-DME) architecture suitable for modern software defined networks. Our main contributions are: (1) we formulate an approach in which measurement matrix in the compressed sensing scheme can be accurately and dynamically estimated through a reformulation of the problem based on traffic demands. (2) We show that the problem formulation using a dynamic measurement matrix based on instantaneous traffic demands may be used instead of a stationary binary routing matrix which is more suitable to modern Software Defined Networks that are constantly evolving in terms of routing by inspection of its Eigen Spectrum using two real world datasets. (3) We also show that linking this compressed measurement matrix dynamically with the measured parameters can lead to acceptable estimation of Origin Destination (OD) Traffic flows with marginally poor results with other state-of-art schemes relying on fixed measurement matrices(4) Furthermore, using this compressed reformulated problem, a new strategy for selection of vantage points for most efficient traffic matrix estimation is also presented through a secondary compression technique based on subset of links . Experimental evaluation of proposed technique using real world datasets Abilene and GEANT shows that the technique is practical to be used in modern software defined networks. Further, the performance of the scheme is compared with recent state of the art techniques proposed in research literature.

## 1 Introduction

Wide Area Network Parameter Monitoring (Network Delays, Traffic Volumes etc) in the Internet also known as Network Tomography has always caught attention from researchers for better

,

network management and future planning [1-3]. Given the scale of the Internet, monitoring all the network is often not feasible; also some of the parameters can be easily observed whereas others cannot. Researchers have formulated several scalable techniques for the estimation of unobservable network parameters borrowing concepts from spatio-temporal methods for statistical signal processing suitable for linear models of wide area networks. However, with the advent of Cloud Computing platforms on the Internet including Infrastructure as a Service (IaaS), Platform as a Service (PaaS) and Software as a Service (SaaS), the traffic patterns have become so diversified that they do not follow one distribution (Gaussian, Poisson, Negative Binomial etc) leading to reduced accuracy of previously proposed methods.

In this work, we revisit the problem of Traffic Matrix estimation expressed through the linear relationship of:

$$Y = AX \qquad (1)$$

In this relationship, $Y \in \Re^{m \times t}$ is the (observable) Link Count Measurements matrix and where $m$ is the number of wide area network links, recorded at time intervals indexed by t. Similarly, $A \in \Re^{m \times n^2}$ is the binary Routing Matrix (or Measurement Matrix) where $m$ is the number of wide area links and $n$ is the number of cloud nodes (resulting in $n^2$ possible paths between cloud nodes). Here, we assume that the routing matrix remains stable (stationary) over long periods of time (unlike wireless networks) which is true for Internet where majority of the wide area paths are stable on the order of several hours or days; $X \in \Re^{n^2 \times t}$ is the Traffic Matrix (unobservable) which is to be estimated through knowledge of the other two at same time intervals indexed by $t$.

Traffic matrix solution to the problem depicted in Equation 1 is often challenging as: (a) *A* is a 'fat' matrix; the system of equations is ill-posed, under constrained and under-determined as $m \ll n$. It is well known fact that an under-determined system of equations may not have a unique solution. In contrast, a well-posed, balanced or over-determined system of equations ($m \geq n$) such that *A* is a 'tall' or a 'square' matrix, may have a unique solution. (b) The assumption of *A* being stationary is no longer valid in this modern age of software defined networks. In this era, routing is not dictated by deterministic routing algorithms but by more 'opportunistic' software defined strategies to makethat allows more user-centric routing decisions rather than network centric. In case of Software Defined Cloud Computing Platforms, some decisions may also be user-centric but network oriented such as to obtain good load balancing on the network servers or network as a whole.

Recent research literature has laid the foundation of compressed monitoring in the field of Network Traffic Matrix Estimation where the above system of equations may be compressed in the form:

$$Y = \Phi X_c \qquad (2)$$

Where, $\Phi = A\Psi$ is a highly compressed measurement matrix in comparison to original measurement matrix $A$, $\psi$ is a sparse representation basis to recover $X_c$, which is the compressed estimation of the most vital components of $X$ to help recompose its accurate reconstruction. Using this new compressed formulation of the problem the initial ill-posed and underconstrained sysem of equations becomes highly regularized.

Our contributions in this work are summarized as follows: (1) we formulate a Compressed Sensing with Dynamic Model Estiamtion (CS-DME) approach in which measurement matrix in the compressed sensing scheme can be accurately and dynamically estimated through a reformulation of the problem based on traffic demands. This compressed formulation of the problem helps us to explore the ideal boundary of the reduced solution space where the original Traffic Matrix Estimation problem switches from under-determined to balanced to over-determined. (2) We show that the problem formulation using a dynamic measurement matrix based on instantaneous traffic demands may be used instead of a stationary binary routing matrix which is more suitable for modern Software Defined Networks that are constantly evolving in terms of routing by inspection of its Eigen Spectrum using two real world datasets. (3) We also show that linking this compressed measurementmatrix dynamically with the measured parameters can lead to accurate estimation of Origin Destination (OD) Traffic flows with results close to other state-of-art schemes relying on fixed measurement matrices (4) Furthermore, using this compressed reformulated problem, a new

,

strategy for selection of vantage points for most efficient traffic matrix estimation is also presented through a secondary compression technique based on selection of a subset of links for monitoring. Experimental evaluation of proposed technique using real world datasets Abilene and GEANT shows that the technique is practical to be used in modern software defined networks. Further, the performance of our scheme is compared with recent state of the art techniques proposed in research literature.The rest of this paper is as follows: Section 2 describes Related Work. Section 3 discusses the proposed approach for compressed sensing of traffic matrix in large software defined cloud networks. Section 4 discusses the real world Internet datasets used in this work. Section 5 presents Performance Evaluation of the proposed scheme and comparison with other algorithms proposed in recent research literature. Section 6 finally concludes the paper.

## 2 Related Work

Estimation of Network parameters for better capacity planning and management is a current topic of research[1-6]. Special terms have been coined in research literature for estimation of network parameters like network delays [4-5] and traffic volumes [1-3]. These techniques have been branded into Kriging [10-12], Cartography [13], Tomography [1-3] and Compressed Sensing [14-18]. Technically, all of three approaches can be divided into space based, time based and spatio-temporal methods [19]. Another notable research contribution by Medina et al. [20] has categorized the first generation state-of-the-art methods as belonging to one of Linear Optimization, Bayesian Estimation, Maximum Likelihood (or Expectation Maximization) approaches. For the problem of traffic matrix estimation, considered in this paper, the latter two approaches work on statistical inference methods on the distribution of traffic matrix elements to compute an expectation of the traffic matrix elements given Link Counts information. However, these techniques rely on an initial assumption of the mean and variance of the traffic flows; for which several researchers have used different distribution: Gaussian, Poisson etc. [6].Even if these initial estimates of the prior distribution of the traffic flows are known, thethesetechniques still heavily depend on the knowledge of the initial prior [20] i.e. the quality of the solution obtained depends on the 'goodness' of this prior solution. Initial work by Tebaldi [21] and Vardi [22] of traffic estimation in IP backbones have proposed that given link load measurements some elements of the traffic matrix may be estimated and the remaining may be computed algebraically, thus reducing the scope of the problem to estimation of smaller number of origin-destination (OD) pair traffic flows. Such schemes may prove beneficial if the routing matrix (A) exhibits a highly decaying Eigen spectrum [11]. Other notable contribution in the field of traffic matrix estimation include, Bayesian Learning techniques for traffic matrix estimation, investigated by Nie [23] and Xiaobo [24].

Other researchers [25-28] have focused their attention on the optimal placement of network monitors for scalable monitoring of the network. Boolean Network Tomography has been a recent topic of interest [29-30]. Qazi and Moors have studied the issue of estimation of network parameters through network tomography when the underlying network model is estimated with errors [31] and Qazi et al. studied the applications of genetic algorithms to the network tomography problem [32]. The problem of the presence of heterogeneous traffic due to the cloud based applications have further increased the complexity of network tomography based estimation techniques [23].Probability Model based Traffic Matrix Estimation technique has been covered for star networks and other general topologies by Tian et al. [33]; However they rely on several assumptions to simplify the model such as fixed routing probability between nodes and do not address, the problem of the implementation of this technique for more general Internet based software defined cloud networks.

,

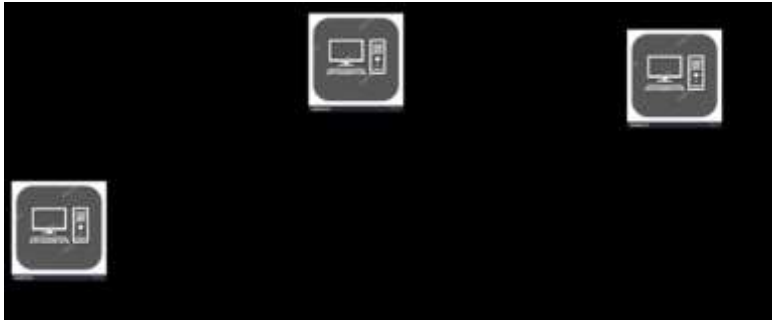

Fig.1A Toy Cloud Network. (Here links are identified by the direction of traffic)

## 3 Proposed Approach for Compressed Sensing of Traffic Matrix based on Traffic Demands

In this section, we describe our approach of compressed sensing. In this formulation, the '$n^2$' paths between the participating '$n$' Cloud Computing Network Nodes $i=1,2,\ldots$, n-1, n are assumed to be sorted as all paths originated from node 1, then all paths originated from node 2 and so on. Henceforth, paths are sorted in the manner: 1-1,1-2,1-3,…, 1-(n-1), 1-n, 2-1,2-2,2-3, …,2-(n-1), 2-n, 3-1, 3-2, 3-3 … etc. The columns of the routing matrix are also assumed to be sorted in this manner for mathematical ease.

It is worth mentioning that real widearea cloud networks measuring traffic origin-destination between the cloud nodes may distinguish between Intra Traffic between the Cloud Nodes and Inter Traffic between the studied Cloud Network and any other outside networks. Thus OD traffic for which source and destination is considered same signifiestraffic sourced at a node and destined for a node in an outside network reachable through the same node while using some links inside the Cloud Network due to routing policies e.g. Hot Potato Routing.

Now, we demonstrate our approach through a toy example of a simple 3 node, 4 link wide area cloud network as shown in Fig.1.Here, the sample Routing Matrix as outlined by the above description will be:

$$A = \begin{bmatrix} 0 & 1 & 1 & 0 & 0 & 0 & 0 & 0 & 0 \\ 0 & 0 & 0 & 1 & 0 & 0 & 1 & 0 & 0 \\ 0 & 0 & 1 & 0 & 0 & 1 & 0 & 0 & 0 \\ 0 & 0 & 0 & 0 & 0 & 0 & 1 & 1 & 0 \end{bmatrix} \quad (3)$$

Next, we define a transformation matrix $\Psi(t)$ based on what fraction of total traffic sourced at each Cloud Nodes is destined towards other Cloud Nodes at each time stamp of the training data. Thus, we define $\Psi(t) \in \Re^{n^2 \times n}$ such that :

$$\Psi_{i,j}(t) = \begin{cases} p_{j,(i-1) \bmod n+1} & \text{if } \lfloor (i-1)/n+1 \rfloor = j \\ 0 & \text{otherwise} \end{cases} \quad (4)$$

,

for every time stamp $t$, where $p_{j,(i-1)\bmod n+1}$ is the fraction (or equivalently probability) of the total traffic of OD flows originating from node $j$ to other cloud nodes. Each row of $\Psi(t)$ can have at most one non-zero, postive value. For our toy cloud network presented in Fig.1, $\Psi(t)$ will become:

$$\Psi(t) = \begin{bmatrix} 0 & 0 & 0 \\ p_{12} & 0 & 0 \\ p_{13} & 0 & 0 \\ 0 & p_{21} & 0 \\ 0 & 0 & 0 \\ 0 & p_{23} & 0 \\ 0 & 0 & p_{31} \\ 0 & 0 & p_{32} \\ 0 & 0 & 0 \end{bmatrix} \quad (5)$$

The Compressed Measurement Matrix $\Phi(t)$ is defined as Probability based Traffic Demand Matrix in the remainder of this paper. It can be obtained directly from this newly defined transformation matrix $\Psi(t)$ and the stationary routing matrix $A$ as:

$$\Phi(t) = A\Psi(t) \quad (6)$$

This new probability based traffic demand matrix $\Phi$ scales the unit entries of the original binary routing matrix $A$ by the probability that a link may be used between a given traffic source in the cloud network source destination pair given the likelihood that there is a traffic between the source destination pair. For our toy cloud network presented in Fig.1 $\Phi(t)$ will be:

$$\Phi(t) = \begin{bmatrix} 1 & 0 & 0 \\ 0 & p_{21} & p_{31} \\ p_{13} & p_{23} & 0 \\ 0 & 0 & 1 \end{bmatrix} \quad (7)$$

In general, $\Phi(t)$ may be rewritten as:

$$\Phi(t) = \begin{bmatrix} p_{1-1*} & p_{1-2*} & p_{1-3*} \\ p_{2-1*} & p_{2-2*} & p_{2-3*} \\ p_{3-1*} & p_{3-2*} & p_{3-3*} \\ p_{4-1*} & p_{4-2*} & p_{4-3*} \end{bmatrix} \quad (8)$$

where $p_{l-n*}$ is the probability that a link $l$ will carry traffic sourced at cloud node $n$. Note that in current scenario these probabilities are also same as the fraction of total traffic from the given source cloud node carried by a particular link.

Thus, the original estimation problem $Y = AX$ with scope of $X$ as estimation of 9 pair-wise OD flows can be reduced to a compressed representation of only 3 Source Traffic Demands $X_c$ from each source:

$$Y(t) = A\Psi(t)X_c(t) \quad (9)$$
$$Y(t) = \Phi(t)X_c(t) \quad (10)$$

with scope of $X_c$ as only estimation of 3 pair-wise OD traffic demands. Furthermore, this relationship is independent of the fact that all links in the network have to be measured. Thus, we can select the most important links in the network revealing the most important information about traffic demands. Suppose, we select a subset $Y_s$ of the links. We can correspondingly select the corresponding rows $\Phi_s(t)$ of $\Phi(t)$.

$$Y_s(t) = \Phi_s(t)X_c(t) \quad (11)$$

The methodology of selecting subset of links will be explained later. The original Traffic Matrix $X$ can be recovered using the relationship:

$$\widehat{X}(t=T) = \widehat{\Psi}(t=T)\widehat{X}_c(t=T) \quad (12)$$

An initial prior solution can be calculated for source node based Traffic Demands can be calculated as:

$$\widehat{X}_{c0} = (\Phi_s^T \Phi_s)^{-1}\Phi_s^T Y_s \quad (13)$$

where $\widehat{X}_{c0}$ is a prior solution for Traffic Demands using the above mentioned relationship. To make sure that the solution also obeys the space based constraints we solve an optimization problem, after supplying our prior $X_{c0}$ as an initial starting point to our optimization problem.

$$MIN_{X_c>0} e(X_c) =$$
$$(Y_s - \Phi_s X_c)^T * (L_{cov})^{-1} * (Y_s - \Phi_s X_c) \quad (14)$$

Subject to:
$$Y_s \leq \Phi_s X_c \quad (15)$$

where $L_{cov}$ is the link covariance matrix(of the links selected for compressed monitoring) which is used as a weighting matrix to give less weights to the error function $e(X)$ resulting from more variable Link Counts. This matrix can be precomputed during the training period to save computational overheads. After solving for the Traffic Demands $X_c$, we can recover the original OD Traffic Matrix using Equation 12.

To estimate the stochastic time varying matrix of Traffic Demands at any time stamp $T_s$, $\widehat{\Psi}(T_s)$ we use training data to develop a relationship between instantaneous link counts of monitored links $Y(t)$ and the traffic demands $\Psi(t)$. $\widehat{\Phi}(T_s)$ can then be estimated using Equation 6. In this work, we assume this is a generalized linear model which gives us satisfactory results. The size of our training data is explained in the next section. Note that since we are only measuring a subset of the links $Y_s(t)$ instead of $Y(t)$, in the initialization phase instead of using $Y_s(t)$ we use the complete $Y(t)$ to make an initial approximation $\widehat{\Phi}(T_s)$ possible; once $\widehat{\Phi}(T_s)$ is estimated we use our initial assumption that only $s$ of $m$ links are monitored.

Now, we describe the methodology of selecting subset of links. Special emphases is given on selection of links according to importance in order to make compressed sensing of the cloud network possible without compromising on estimation of OD flows. Particularly,we use the subset

,

selection method based on Algorithm 12.2.1 presented in [34]. We first carry out Singular Value Decomposition of compressed probability based Traffic Demand matrix $\Phi$ of the cumulative OD flows sourced from each source.

$$svd(\Phi) = U^T S V \quad (16)$$

where, $S$ is the diagonal matrix revealing the Singular Values and $U$ and $V$ are the left and right singular vectors respectively. In the next step, we carry out qr-decomposition with column pivoting of the orthogonal basis of the compressed Traffic Demand matrix $\Phi$ through selection of $k$ rightmost columns of the matrix $U$ which we refer to as $U_k$. This process selects a subset of the $k$ rows of the compressed Traffic Demand matrix $\Phi$ corresponding to the $k$ most important links.

$$qr(U_k^T) = q \, r \, P \quad (17)$$

since, we want to select the rows (instead of columns) of the compressed Traffic Demand matrix $\Phi$ corresponding to the links that we wish to monitor.

Finally, the permutation step in accordance with the permutation matrix $P$ gives us the selected links $Y_s$ and the selected rows Traffic Demand matrix $\Phi_s$.

$$Y_s = first \, k \, rows \, of \, P^T Y \quad (18)$$

$$\Phi_s = first \, k \, rows \, of \, P^T \Phi \quad (19)$$

We conclude this section by presenting the pseudo-code of our complete scheme i.e. CS-DME as Algorithm 1.

Another important point of worth highlighting is the chicken and egg problem in selection of a subset of links based on an already dynamic compressed measurement matrix $\Phi$ based on Traffic Demands. In this work, we assume that $\Phi(t)$ is entirely depending on the measured links in the network deemed to influence the estimation of traffic matrix. Thus, the link selection may change over time based on evolution of $\Phi$ in our model. This can also be observed in Algorithm 1.

## 4 Datasets

We use two different real world data sets to evaluate the peformance of our proposed algorithm, Compressed Sensing through Dynamic Model Estimation (CS-DME) for traffic matrix estimation in software defined cloud networks. The motive of our analysis is to assess the ability of our approach as comprehensively as possible in a real world scenario. To that end, we use two data sets namely Abilene [13] and GÉANT [14]. Each of the data sets is described in detail.

,

**Algorithm 1.** Pseudocode of the Compressed Sensing algorithm CS - DME.

**Given :** *For a cloud network with n nodes and m links*

*Traffic Matrix* $X(t) \in \Re^{n^2 \times t}$, *Routing Matrix* $A = [0,1]^{m \times n^2}$ & *Link Counts* $Y(t) \in \Re^{m \times t}$

**Training Phase:**

**Input :** $X(t), Y(t) ; t = 1, 2, \ldots, |training\ data\ size|$

*Using $X(t)$, Compute* $\Psi(t) \in \Re^{n^2 \times n}$

*where* $\Psi_{ij} = \dfrac{Traffic\ Demand\ X_{j,(i-1) \bmod n+1}}{Total\ Traffic\ Demand\ X_{j,*}}$ *if* $\left\lfloor \dfrac{(i-1)}{n} \right\rfloor + 1 = j$, *otherwise* $\Psi_{ij} = 0$

**Output :** *Robust regression coefficient matrix* $\beta \in \Re^{(m+1) \times n^2}$ *for each row - wise positive element of $\Psi(t)$ & $Y(t)$. Each row of $\Psi(t)$ can have at most one postive element.*

**Estimation Phase (on Test Data)**

*Initialize* $Y_r(t-1) = Y_r(t)$, $P = Identity\ Matrix\ of\ size\ m$

*for t = start of testing interval:end of testing interval*

**Given :** $Y_s(t) = monitored\ links$, $\widehat{Y}_r(t-1) = Unmonitored\ links$, $\beta$, $subset\ links\ s = size(Y_s(t))$

    *Initialize $\Psi(t)$ as all zeros.*

$$\widehat{Y}(t) = (P^T)^{-1}[Y_s(t); \widehat{Y}_r(t-1);]$$

    *Estimate $\Psi(t)$ as follows :*

    *for* $i = 1:1:n^2$

      *for* $j = 1:1:n$

        *if* $\left\lfloor \dfrac{(i-1)}{n} \right\rfloor + 1 = j$

          $\widehat{\Psi}_{ij}(t) = \beta_{:,i}^T \times [1; \widehat{Y}(t)];$

          *break;*

        *endif*

      *endfor*

    *endfor*

    *Estimate* $\widehat{\Phi}(t) = A \times \widehat{\Psi}(t)$

    *Selecting a Sub - set of links $Y_s$ to monitor:*

    $USV^T = SVD(\Phi)$

    $QR = U_k^T P$    $(U_k = U(:,1:k))$

    $Y_s = P^T \widehat{Y}(t)$ & $\Phi_s = P^T \widehat{\Phi}(t); L_{cov} \in \Re^{k \times k} = Cov(Y_s) (pre$ - $computed\ in\ training\ period))$

    *Solve the optimization problem :*

$$\underset{X_c \geq 0}{\arg\min} (Y_s - \Phi_s X_c)(L_{cov})^{-1}(Y_s - \Phi_s X_c)^T \quad subject\ to: \quad Y_s = \Phi_s X_c$$

    *Finally,* $\widehat{X}(t) = \Psi(t)\widehat{X}_c(t)$ & $\widehat{Y}(t) = \Phi(t)\widehat{X}_c(t) = (P^T)^{-1}[\widehat{Y}_s(t); \widehat{Y}_r(t);]$

*endfor*

,

### 4.1 Abilene Dataset

The Abilene Dataset (aka Internet2 since 2007) provides traffic measurement data and routing information for a high performance Backbone network in the US. The dataset provides a known Routing Matrix and recorded values of Origin-Destination (OD) Traffic counts (TM) for a consecutive 3 week period on the Abilene network with 11 WAN nodes and 41 directed WAN links. Thus the Abilene network carries a cumulative traffic for a total of 121 possible Origin-Destination (OD) flows. Link Counts can be easily calculated through the relationship given by Equation 1. The OD byte counts are sampled into 05 minute time intervals, resulting in 2016 byte count samples for each week. Same numbers of samples are thus also available for Link Counts data across each of the 41 links.

### 4.2 GÉANT Dataset

GÉANT is an experimental network for the education and research community similar to Abilene but based in Europe. This dataset provides anonymized topology information and OD Traffic Data (TM) and for a consecutive 4 month period for a total of 23 WAN nodes and 74 directed WAN links. Thus the GÉANT network carries a cumulative traffic for a total of 529 possible Origin-Destination (OD) flows. This dataset provides traffic matrices (TM) in a XML format which are measured on 15 minutes time interval. We developed a MATLAB based XML parser to obtain the TM and the routing matrix; Link Counts are then obtained in a similar manner as with Abilene Dataset using Equation 1.

In both the datasets, we use 2000 consecutive time samples of which the first 500 and 1500 are used as training dataset to predict traffic demands in Abilene and GÉANT dataset respectively. These are required for populating the entries of $\Psi$ as described in Section 3. The reason for using large training dataset is that we found several extended periods that have no recorded OD flows for certain source destination pairs in both datasets. This could be due to missing measurements.

## 5 Performance Evaluation

First, we carry out an intuitive analysis of our proposed approach using real world datasets representative of traffic on wide area cloud networks. The purpose of this analysis, is to validate if our spatial compression of the $A$ and $X$ matrix decreases the estimation efficiency of the original problem. For that we compare the Eigen Spectrum of the network observation matrix in both cases. Fig.2 shows graphs depicting the Eigen spectrum of the original Binary routing matrix and spatially compressed Traffic Demands matrix. For the Abilene Dataset, the dimensions for original binary routing matrix and spatially reduced compressed Traffic Demand probability matrix are 41 x 121 and 41 x 11 respectively. For the GÉANT dataset, the dimensions of the original binaryrouting matrix and the compressed Traffic Demand probability matrix are $74 \times 529$ and $74 \times 23$. None of these matrices are rank deficient and the rank is denoted by the smaller of the two figures.

,

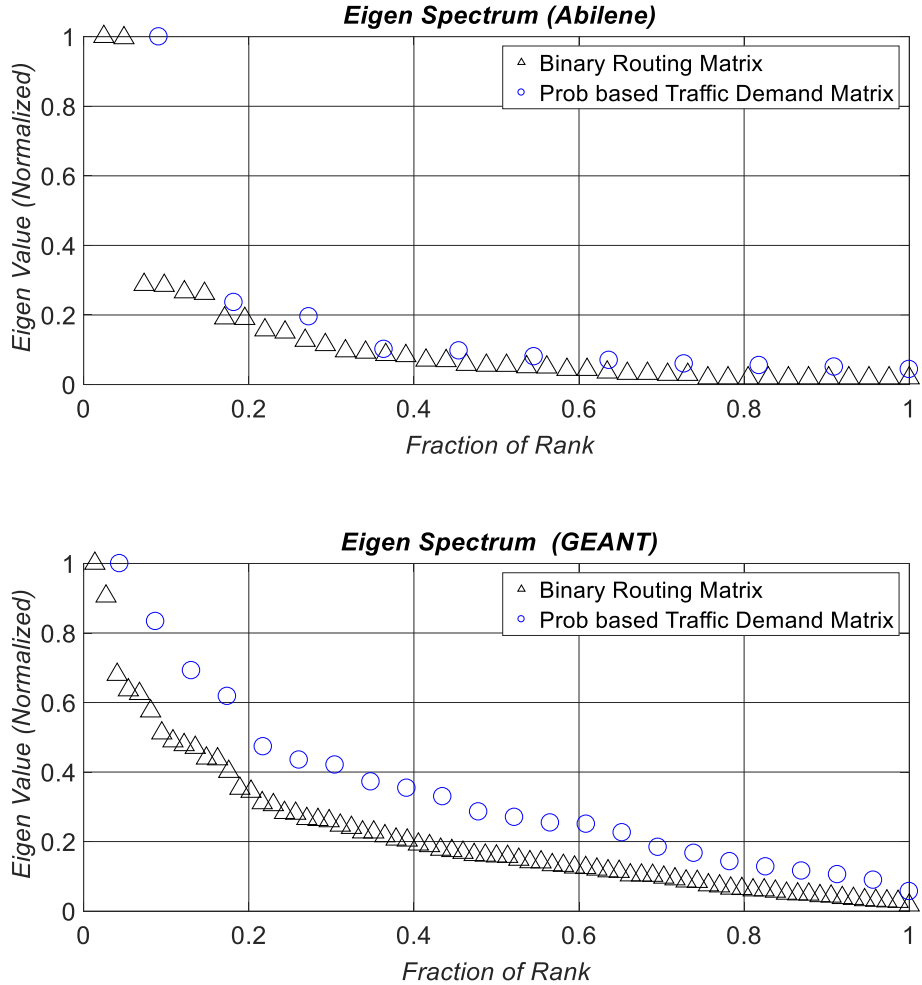

Fig. 2 Retention of Compressibility: Comparison between with the compressed dynamic traffic demand matrix Vs original stationary routing Matrix

The decay in the Eigen values roughly signifies the amount of information that can be extracted for the unknowns by solving a sub-set of the linear system of equations. We notice that the Eigen spectrum of Abilene shows sharper decay than that of GÉANT. Further, we observe a greater difference between the Eigen Spectra in the GÉANT network than Abilene. This shows that Abilene retains its compressibility characteristics better than GÉANT. However, a general trend of decay of the normalized Eigen spectrumfor both Abilene and GÉANT network across both networks suggests retention of compressibility characteristics of the original set of equations even when the probability based compressed Traffic Demand probability matrix $\Phi$ is used instead of traditional routing matrix $A$.

In the Abilene dataset, we observe that the decay of Eigen values corresponding to 6 link measurements in the original system of equations as indicated by Equation 1 is roughly equivalent to the monitoring of just 2 links in the Traffic Demands based reduced scope of original problem (Equation 11). Monitoring all 41 links in the old systems is now reduced to just monitoring of 11 links as the new system of equations becomes balanced at this point as explained earlier. Similar observations can be made for GÉANT; Monitoring all 74 links in the original problem is now just reduced to the problem of monitoring just 23 links. However, in practice the number of links may be higher than these figures. This is due to the fact that our compressed measurement matrix is dynamic and noise is added to the compressed measurement matrix when a subset of the links are monitored.

Fig.3 sheds more light on this observation. Even when the Network Routing matrix is spatially compressed based on Traffic Demands rather than individual OD traffic flows, the matrix retains sufficient orthogonality as shown by the spiky surface on the 3D plots of the Traffic Demand probability matrix. This permits further compressed sensing using only a subset of link

,

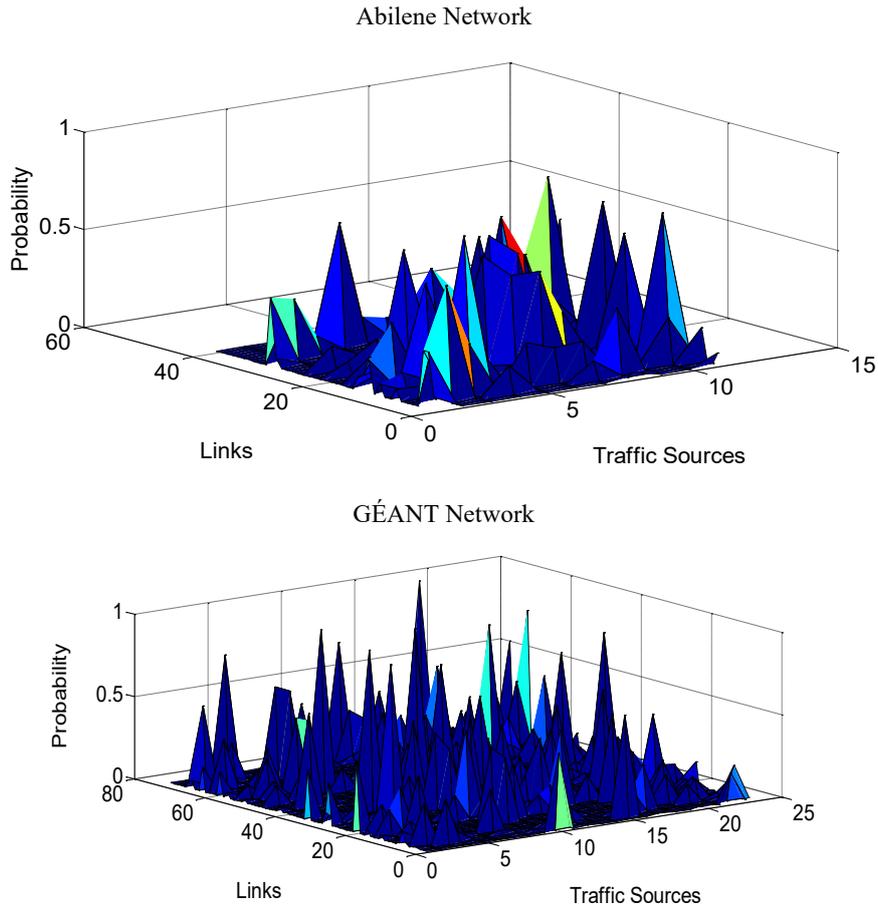

Fig. 3 Retention of orthogonality in the compressed measurement matrix of Abilene and GÉANT Networks.

measurements. Here, the z dimension shows the probability that a particular link will be used by a particular Traffic Source, peaks at or near the value of unity demonstrate links exclusivity to a particular Traffic Source. In Fig.2, we observe that although the Eigen values of the Traffic Demands matrix of both Abilene and GÉANT shows exponential decay, still GÉANT shows more smoother decay with respect to Eigen values. Another related aspect of this is the fact that there is more link sharing between OD pairs in the Abilene compared to GÉANT networks. This would mean more stability of the Transformation Matrix $\Psi$ (Equation 6) which is used to convert the solution of Source Traffic demands into OD traffic volumes, our goal.

Fig. 4 and Fig. 5 depict the results of using the proposed scheme for Abilene and GÉANT networks when variable number of link measurements (link counts) are selected and the amount of error in the estimation of OD traffic flows. As monitored paths are increased the OD estimation error is reduced. Fig. 6 and Fig. 7 show the tracking efficiency of our algorithm for two sample OD flows each in Abilene and GEANT networks. We find that OD flows in Abilene network are best predicted when at least 35 of the 41 links are monitored. For GEANT good OD tracking efficiency is possible when at least 65 of 71 paths are monitored.

Now, we compare the performance of our proposed scheme (CS-DME) with three other state-of-the-art Compressed Sensing (CS) schemes proposed in recent research literature; Principal Component Analysis (PCA) proposed by Soule et al. [18], CUR Decomposition (CUR) proposed by Kumar et al. [17] and Probability based Model Estimation, proposed by

,

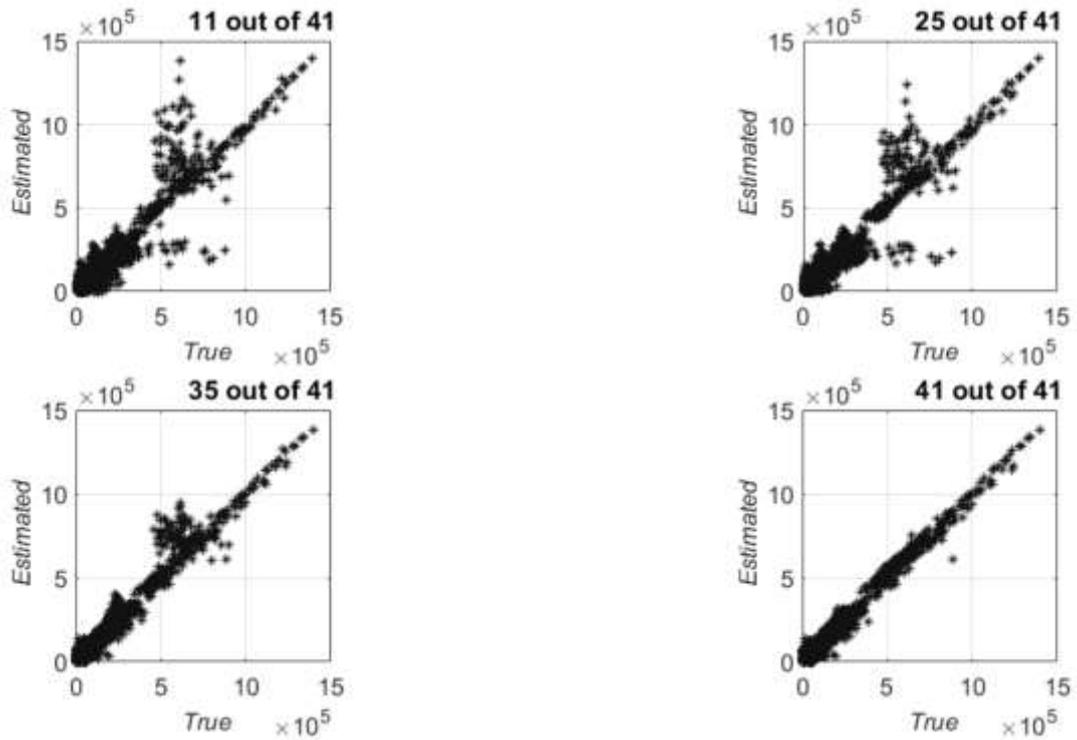

Fig.4 Abilene Network when varying number of links (out of 41) are measured mentioned on top of each figure. Prediction made after using training data of first 500 time stamps. Each time step is 5 minutes in Abilene Dataset and the units of OD flows are in kbps

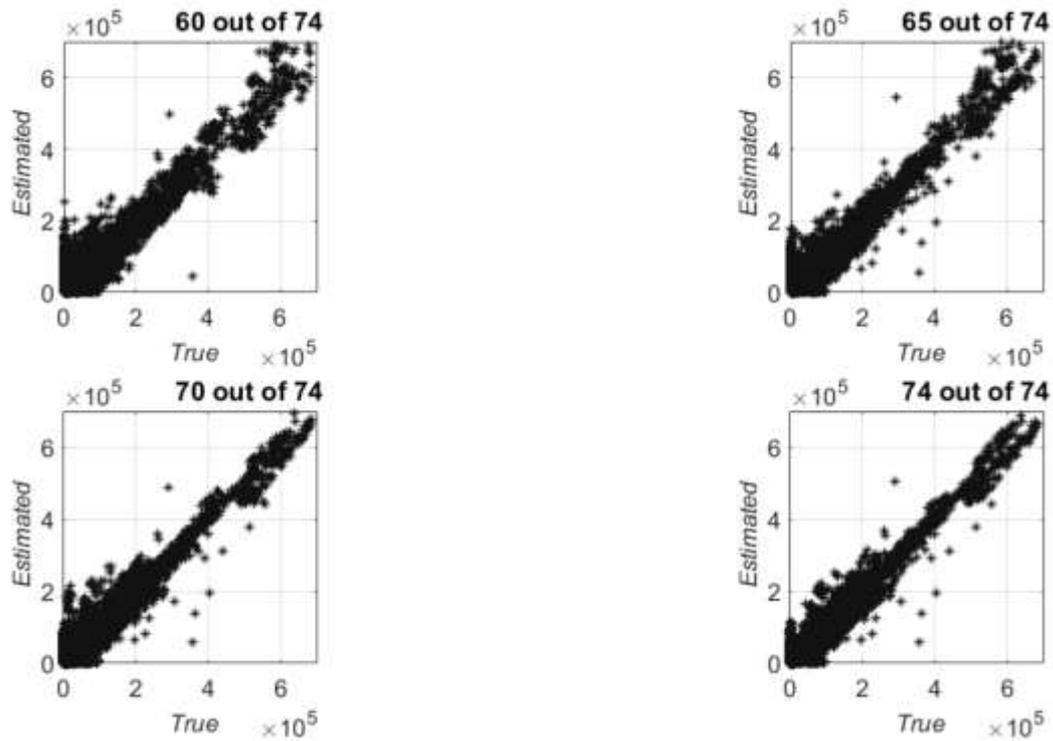

Fig.5 GEANT Network when varying number of links (out of 74) are measured mentioned on top of each figure. Prediction made after using training data of first 1500 time stamps. Each time step is 15 minutes in GEANT Dataset and the units of OD flows are in kbps

,

Tian et al. [33] which we refer to as CS-PCA, CS-CUR and CS-PME respectively in the remainder of this paper. The first two schemes (CS-PCA & CS-CUR) are state of the art compressed sensing schemes which differ from our scheme in the fact that although they allow for the measurement matrix to be compressed efficiently using $k$ principle components they do not allow for a dynamic measurement matrix in the model such as one based on Traffic Demands suitable for modern software defined networks. Similarly, CS-PME considers a measurement matrix based on Traffic Demands but assumes that these Traffic Demands are fixed or well known to within small errors instead of variable.

The other compressed schemes reduce the dimensions of the measurement matrix just like our scheme but with the following notable differences. The implementation of the first two techniques are based on the PCA and CUR decomposition of the matrix $AX(t)$ to compute the static compressed measurement matrix $\Phi$ where $X(t)$ is considered during the training interval as described above. Thereafter, any $k$ number of components corresponding to OD traffic flows may be selected from this decomposition to formulate the compressed measurement matrix. For the implementation of the third technique CS-PME inspired from Tian's algorithm [33], we use the compressed measurement matrix $\Phi$ formulated using the probabilistic Traffic Demands with the difference that original Tian's algorithm assumes a prior noisy model of traffic demands to estimate traffic demands in real time. Thus, in the implementation of Tian's scheme to consider the affects of noise we randomly pick a Traffic Demand value from the Normal (Gaussian) pdf $N(\mu, \sigma^2)$ with μ=OD flow mean and σ= 0.4 times the OD flow mean as obtained from the training data for computation of $\Phi$.

The justification for the selection of this noise model isdue to the relationship between the OD means and variance in the Abilene and GEANT dataset. Fig. 8 shows the relationship between mean and variance of the OD Flows in Abilene and GEANT Network. From these plots, it is clear that OD traffic volumes show very high variability; the variance values are five orders of magnitude larger than the mean values for elephant OD flows and 3 orders of magnitude larger than the mean values for mice OD flows in Abilene Dataset.The coresponding values are 4 and 2 for the GEANT dataset. Thus, using the noise model above CS-PME will be able to predict overall OD flows with better prediction of mouse flows than the other three schemes. Its peformance will be better for GEANT as GEANT has a higher proportion of mouse flows compared to Abilene as observed from Fig. 8.

The value of $k$ is tunable for the compressed schemes CS-PCA and CS-CUR where $k$ can beselected anywhere between 1 and $n^2$, i.e. total number of OD flows to be estimated. However, the value of $k$ is fixed to the number of nodes $n$ in Abilene / GEANT for CS-DME and CS-PME as in these two schemes we consider OD Demands of all traffic sources instead of all individual OD traffic flows. Thus the same value of $k$ is used for a fair comparison in the performance of all four algorithms in the remaining results shown in this paper i.e . $k=11$ for Abilene and $k=23$ for GEANT. None of the three schemes (CS-PCA, CS-CUR and CS-PME) assumes a real-time dynamic model between the measured parameters and the compressed measurement matrix for estimation of the latter. This is where our scheme differs when doing the comparison with all three algorithms as owing to this dynamic model, our scheme can further reduce the number of links that need to be monitored at a particular time in the network. As evident from Fig. 6 -7, we select the number of links used for monitoring to be 35 and 65 in Abilene and GEANT respectively leading to approximately 14% reduction in link monitoring overheads for both Abilene and GEANT. The reduction may be increased further if only anomalies are to be detected at the cost of increased prediction error. We leave this as future work.

,

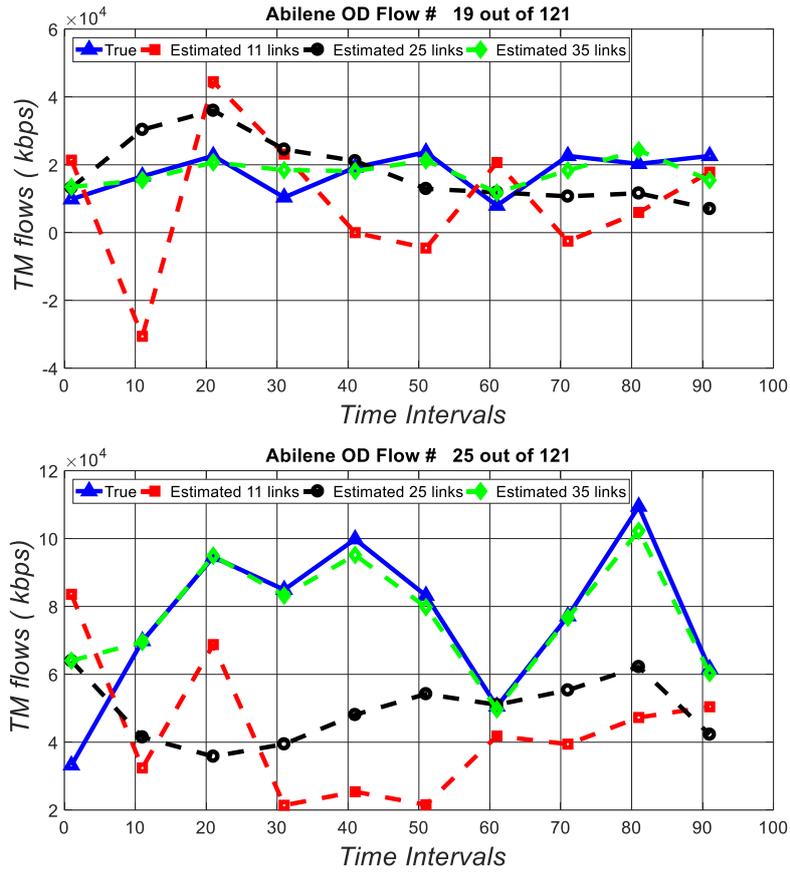

Fig. 6 Abilene Network when varying number of links (out of 41) are measured. Prediction made after using training data of first 500 time stamps. Each time step is 5 minutes in Abilene Dataset.

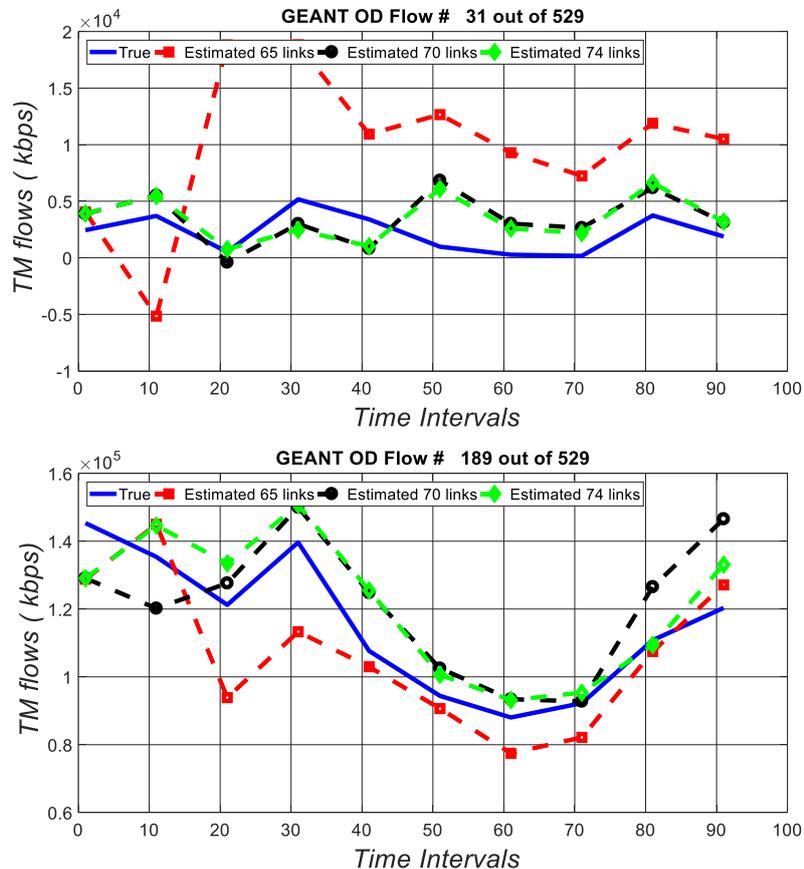

Fig. 7 GEANT Network when varying number of links (out of 74) are measured. Prediction made after using training data of first 1500 time stamps. Each time step is 15 minutes in GEANT Dataset.

To compare the performance of our scheme against other compressed sensing schemes for estimation of Traffic Matrix, we use the metrics of Spatial Relative Error (SRE), Temporal Relative Error (TRE), Bias and Standard Deviation used widely in recent research literature [35].

Spatial Relative Error (SRE) is defined as:

$$SRE(i) = \frac{\|\widehat{X}_{i,t} - X_{i,t}\|_2}{\|X_{i,t}\|_2} \quad (20)$$

Temporal Relative Error is defined as:

$$TRE(t) = \frac{\|\widehat{X}_{i,t} - X_{i,t}\|_2}{\|X_{i,t}\|_2} \quad (21)$$

Bias is defined as:

$$Bias(i) = \frac{1}{T}\sum_{t=1}^{T}(\widehat{X}_{i,t} - X_{i,t}) \quad (22)$$

Standard Deviation is defined as:

$$Standard\ Deviation(i) = \sqrt{\frac{1}{T-1}\sum_{t=1}^{T}((\widehat{X}_{i,t} - X_{i,t}) - bias(i))^2} \quad (23)$$

where $X_{i,t}$ and $\widehat{X}_{i,t}$ represent the actual and the estimated OD flow $i \in 1,2,\ldots,n^2$ at time instant $t$ and $T$ is the size of the testing period.

Fig. 9 shows the CDF of the SRE of our scheme compared with three other state-of-the-art Compressed Sensing (CS) schemes as described above. From Fig 9 we observe, that due to the dynamic estimation feature of CS-DME, our scheme has a performance superior to CS-CUR and almost similar performance CS-PCA even if less number of paths are monitored (41&74 in CS-PCA & CS-CUR compared to 35& 65 in our schemefor Abilene and GEANT respectively). For the Abilene and GEANT networks, our scheme has spatial relative error of less than 0.8 for approximately 90% and 60% of the OD flows. CUR decomposition technique is well known to be less accurate than PCA in computation of the principal components of the compressed sensing scheme. Interestingly, comparison with CS-PME serves as an asymptotic upper bound for our scheme if the traffic demands are ideally known. Thus, we expect CS-PME to perform better in the prediction of the mouse OD flows with smaller means thus resulting in smaller standard deviation in our noisy prediction model with better Traffic Demands prediction possible for CS-PME. Thus, the upper band of the cdf with typically higher SRE values for mouse OD flows show expectedly worse behavior in the other three schemes (CS-PCA, CS-CUR and CS-DME) compared to CS-PME. This trend is reversed for elephant OD flowswith higher means where CS-PCA and CS-DME outperform all others. The same trend is seen in the CDF of the SRE for GEANT in the performance of CS-PCA, CS-CUR and CS-DME algorithms with improved performance for CS-PME mainly due to the higher proportion of mouse OD flows in the GEANT network as evidenced by Fig. 8.

Our corresponding results for the Temporal Relative Error (Fig. 10) show minor performance degradation in performance of CS-DME. For Abilene, a TRE of 0.3 or less in 80% of the cases for CS-DME compared to a figure of 0.15 and 0.2 for CS-PCA and CS-CUR respectively even when less number of paths are measured in our scheme, 35 in CS-DME vs 41 in CS-CUR & CS-PCA. This result is expected as in our second stage of compression the sub-set selection of links for monitoring is based on spatial considerations so slight degradation is expected in the temporal domain. CS-PME with corresponding figure of 0.55 performs considerably poorly to the other three schemes; this mainly arises again due to constant probability model for OD flows as explained earlier, it is only able to well predict the mouse OD flows; in Abilene the proportion of mouse OD flows is less. For GEANT, CS-DME performs exceptionally well with a TRE of less than 0.2 for 95% of the cases, again close behind CS-PCA and CS-CUR when 15% less link counts are

,

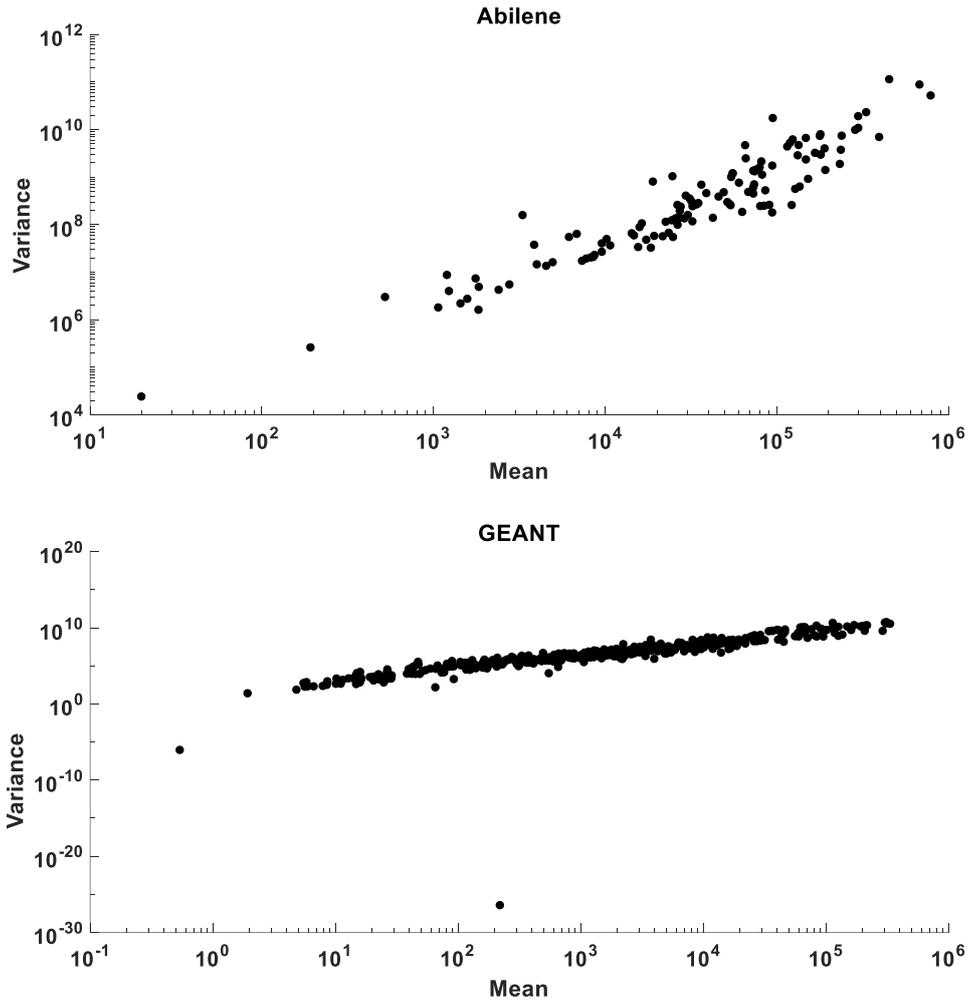

Fig. 8 Relationship between Mean and Variance of OD Flows for Abilene and GEANT. Note the logarithmic scale of both axes

considered in CS-DME. The performance of CS-PME is slightly higher than in Abilene due to higher proportion of mouse flows that are predicted well by CS-PME. Both results of SRE and TRE, provide substantial evidence that our scheme can provide good performance bounds for software defined networks in spatial and temporal domains where the compressed measurement matrix may be dynamic instead of static and lesser number of links may be monitored.

Fig. 11 shows the statistical measure of Bias Vs OD flows sorted in descending order of mean for Abilene and GEANT network. As observed in the results of Spatial Relative Error and Temporal Relative Error, CS-PME has a superior knowledge of mouse OD flows with small means, thus bias values are low for these OD flows with bias values increasing for OD flows with larger means. CS-DME performs closest to CS-PCA with less number of paths being measured in Abilene but slightly worse in GEANT owing to higher noise due to higher compression of the measurement matrix. Fig. 12 shows the statistical measure of Bias Vs standard deviation. Again this figure shows CS-DME has lower standard deviation of error and performance closest to CS-PCA even with 15% lower monitoring overheads in Abilene and GEANT. CS-PME has highest standard deviation mainly due to the ill estimated elephant OD flows compared to all other 3 schemes.

,

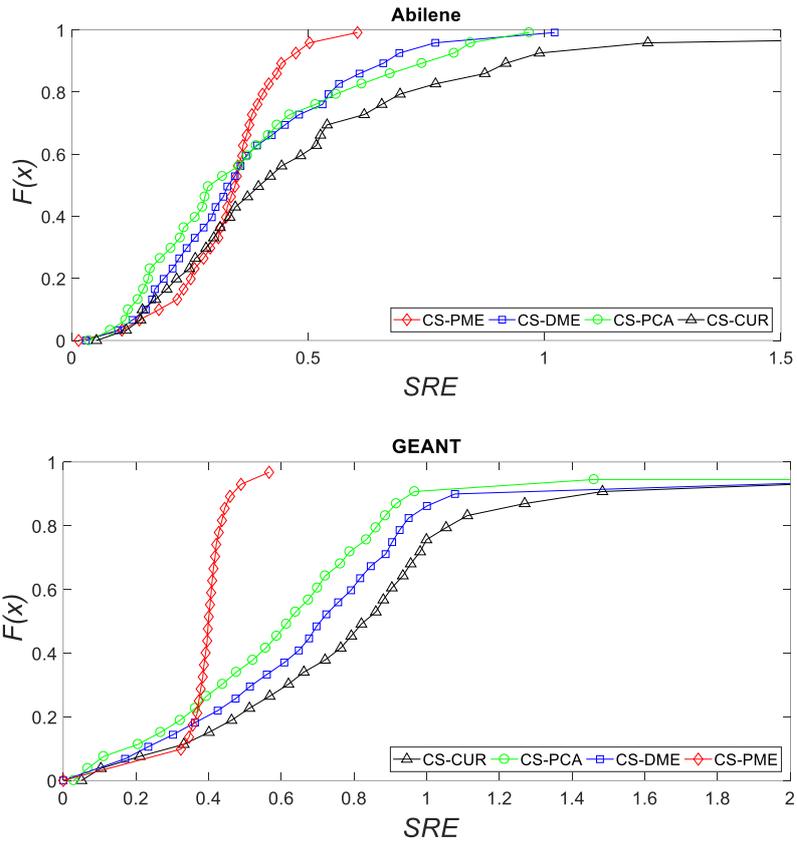

**Fig. 9** Comparison of the CDF of Spatial Relative Error (SRE) for Abilene and GEANT.

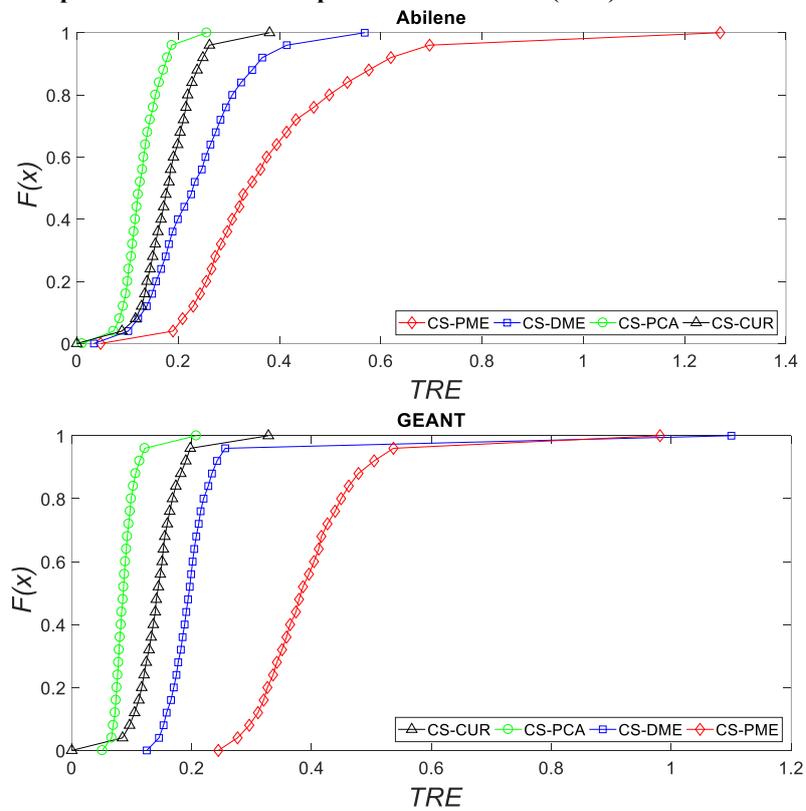

**Fig. 10** Comparison of the CDF of Temporal Relative Error (TRE) for Abilene and GEANT.

,

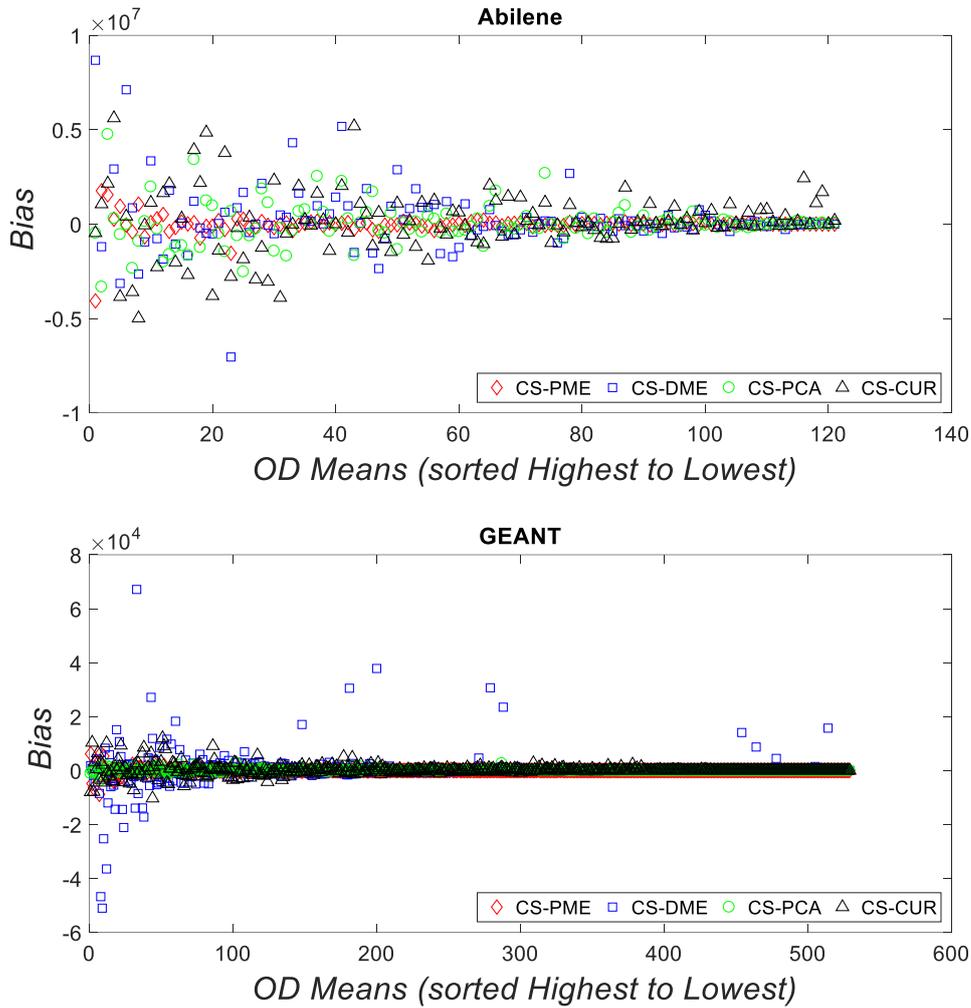
**Fig. 11** Plot of Bias Vs OD Flow Means sorted in descending order for Abilene and GEANT

## 6 Conclusion

In this paper, we presented a novel approach for compressed sensing for the traffic matrix estimation for Internet scale, wide area Software Defined Cloud Networks. The presented scheme showed a dynamic model for compresed sensing at reduced network monitoring overheads. This scheme is readily adaptable to the dynamic traffic routing models prevalent in Software Defined Cloud Networks. With up to 15% reduction in link monitoring overheads, OD flows can still be estimated with good tolerance. Comparison with other state-of-the-art schemes shows comparable level of performance against all benchmark tests.

## References

,

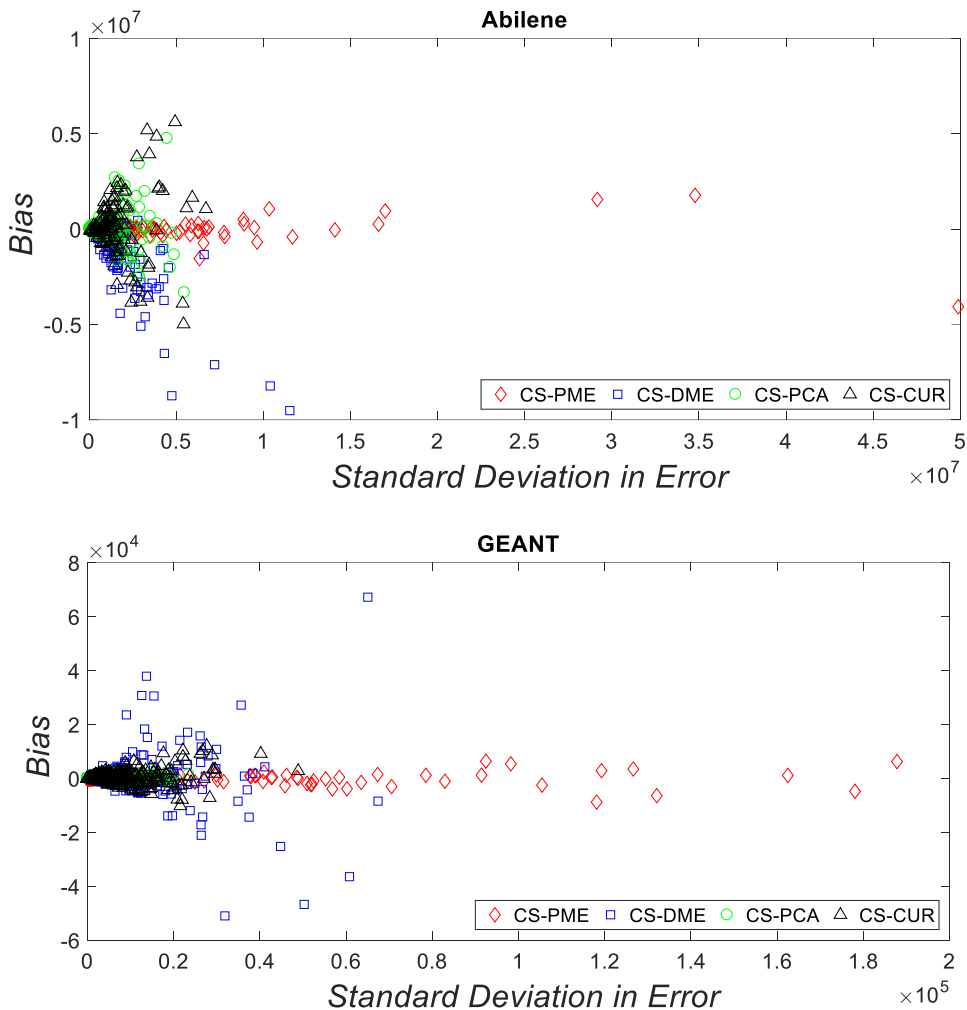

Fig. 12 Plot of Bias Vs Standard Deviation of Error for Abilene and GEANT

,

,